\documentclass{ifacconf}

\usepackage{graphicx}      
\usepackage{subcaption}
\usepackage{natbib}        
\usepackage{amssymb}
\usepackage{amsmath}
\usepackage{amsfonts}  
\usepackage{algorithm}
\usepackage{algorithmic}
\usepackage{enumitem}
\begin{document}
\begin{frontmatter}

\title{Soft Switching Expert Policies\\
for Controlling Systems \\
with Uncertain Parameters} 

\thanks[footnoteinfo]{This work was partially supported by JSPS KAKENHI Grant Numbers 26K21173.}

\author[First]{Junya Ikemoto} 

\address[First]{Graduate School of Engineering The University of Osaka,\\ 
   2-1 Yamadaoka, Suita, Osaka, Japan \\(e-mail: ikemoto@eei.eng.osaka-u.ac.jp)}

\begin{abstract}                
This paper proposes a simulation-based reinforcement learning algorithm for controlling systems with uncertain and varying system parameters. While simulators are useful for safely learning control policies, the reality gap remains a major challenge. To alleviate this challenge, we propose a two-stage algorithm. First, multiple control policies are learned for systems with different system parameters in a simulator. Second, for a real system, the control policies are adaptively switched using an online convex optimization algorithm based on observations. This approach is expected to reduce learning complexity compared with existing approaches that rely on a single policy to address the reality gap.
\end{abstract}


\end{frontmatter}

\section{Introduction}
{\it Reinforcement learning} (RL) is a machine learning framework  that addresses sequential decision-making problems \citep{Sutton_RL}. In RL, an \it agent \rm collects experience data through interactions with a system and automatically learns a control policy based on the collected data. Recently, {\it deep RL} (DRL) has attracted much attention as an effective framework for controlling complex systems. Its applications are wide-ranging, including robot manipulation \citep{RL_manipulation}, autonomous driving \citep{RL_autonomous}, and industrial plant operations \citep{RL_plant}. These successes can be attributed to the ability of {\it deep neural networks} (DNNs) to effectively handle high-dimensional data and accurately approximate nonlinear functions. However, DRL requires a large amount of data through repeated interactions with systems to fully exploit its potential. For physical systems, data collection tends to be costly, and almost random interactions during the early stages of learning can cause catastrophic equipment damage. In such cases, simulators are commonly used. Simulators are valuable for leveraging the capability of DRL to control physical systems because they can be easily accelerated and parallelized to safely collect a large amount of data \citep{Sim_RL}. 

Although simulators provide substantial advantages for DRL, learning in simulators faces a major challenge of bridging the {\it reality gap}, which refers to the mismatch in behaviors between simulated and real systems. If a control policy is learned for a simulated system that poorly represents the real system, it may not perform as expected. Even if a high-fidelity simulator is available, it remains challenging to accurately identify the {\it system parameters} of the real system (e.g., mass and friction coefficients). To learn control policies that are robust to the reality gap, {\it domain randomization} (DR) has been proposed \citep{DR,understanding_DR}. In DRL with DR, the uncertain system parameters are randomized within a given range during the learning phase in a simulator instead of being fixed, which exposes the agent to diverse system dynamics. Despite its conceptual simplicity, DR has successfully bridged the reality gap in complex robotic systems \citep{DR}. Nevertheless, a control policy learned with DR needs the system parameters of a real system to determine control actions appropriately. When the system parameters are not directly observable, recurrent architectures are often employed to infer the latent information from observation histories. However, methods with recurrent architectures typically require specialized training procedures to handle temporal dependencies in partially observable settings, which may increase training complexity \citep{Reccurent_DRL}. 
Additionally, the effectiveness of DR may become limited when the parameter space is large and the corresponding system behaviors differ significantly across parameters. Since DR optimizes an expected objective over randomly sampled system parameters, learning a uniformly effective policy over the entire parameter space can become increasingly challenging. Therefore, instead of learning a single policy over the entire parameter space, we construct multiple expert policies offline and adaptively switch among them online through a lightweight adaptation mechanism.

In this study, we focus on systems whose physical dynamics structure is known while some system parameters remain uncertain or time-varying. To handle parameter uncertainty without recurrent architectures, we propose a two-stage learning algorithm.
In the first stage, multiple representative points are heuristically selected from a predefined parameter range, called the {\it system parameter space}. For each representative point, an {\it expert policy} is learned for the corresponding simulated system using a standard DRL algorithm, such as {\it deep deterministic policy gradient} (DDPG) \citep{DDPG}. In the second stage, we construct a high-level policy, which is called an {\it adaptive policy}, as a convex combination of the expert policies, where the weights are adjusted online based on observations from a real system, thereby enabling smooth transitions among the expert policies. Specifically, we estimate a {\it similarity vector}, which quantifies how accurately each representative simulated system predicts the observed transition of the real system, using {\it online convex optimization} (OCO) \citep{OCO_text_book}, and use it directly as the weights in the adaptive policy. OCO is a well-established framework for sequential online decision-making under uncertainty and enables lightweight online adaptation. To this end, we formulate a convex loss function based on observations from the real system. 

The paper is organized as follows: Section 2 formulates the problem. Section 3 briefly reviews the fundamentals. Section 4 presents a two-stage learning algorithm with a simulator. Section 5 demonstrates the results of the proposed algorithm. Finally, Section 6 concludes the paper and discusses future work.

\it Notation: \rm 
$\mathbb{R}$ denotes the set of real numbers. $\mathbb{R}_{\ge0}$ denotes the set of nonnegative real numbers. $\mathbb{R}^{n}$ denotes the $n$-dimensional Euclidean space. $E[\cdot]$ denotes the expectation operator. 

\section{Problem Formulation}
We consider a discrete-time nonlinear system governed by 
\begin{eqnarray}
x_{t+1}=f(x_{t},a_{t};\xi_{t}),\label{dynamical_system}
\end{eqnarray}
where $x_{t}\in\mathcal{X}\subseteq\mathbb{R}^{n_x}$ and $a_{t}\in\mathcal{A}\subseteq\mathbb{R}^{n_a}$ denote the state and  control input at time $t=1,2,...$, respectively. Let $\mathcal{X}$ and $\mathcal{A}$ denote the state and control input spaces, respectively.  $\xi_{t}\in\Xi\subseteq\mathbb{R}^{n_{\xi}}$ is the vector of system parameters at time $t$, where $\Xi$ denotes the system parameter space. $f:\mathcal{X}\times\mathcal{A}\times\Xi\to\mathcal{X}$ denotes the nonlinear dynamics of the system. The initial state $x_1$ is drawn from a probability density function $\rho_1:\mathcal{X}\to\mathbb{R}_{\ge0}$. 

We emphasize that this study assumes:
\begin{enumerate}[label=(\roman*)]
    \item the dynamics structure $f$ is known,
    \item uncertainty appears only in possibly time-varying system parameters $\xi_t$,
    \item the parameter range $\Xi$ is known beforehand, and
    \item full state observations are available without observation noise.
\end{enumerate}
Since the dynamics structure $f$ is known, the proposed approach is not intended for fully black-box systems and can exploit structural knowledge of $f$ during offline training in a simulator. However, uncertainty in system parameters $\xi_t$ can still lead to discrepancies between simulated and real systems, giving rise to the reality gap. Ideally, we seek a control policy $\mu:\mathcal{X}\times\Xi\to\mathcal{A}$ that maximizes long-term cumulative rewards: $\sum_{t=1}^{\infty}\gamma^{t}R(x_t,\mu(x_t,\xi_t))$, where $\gamma\in(0,1)$ is a discount factor and $R:\mathcal{X}\times\mathcal{A}\to\mathbb{R}$ is an immediate reward function. However, accurately estimating the system parameter $\xi_t$ online may be difficult during control operation. Therefore, instead of assuming direct access to $\xi_t$, we aim to learn a state-feedback control policy that achieves acceptable long-term performance for as wide a range of system parameters as possible.

\section{Preliminaries}
\subsection{Reinforcement Learning (RL)}
An RL problem is formulated by a {\it Markov decision process} (MDP) $\mathcal{M}=\left<\mathcal{X},\mathcal{A},R,p_f,\rho_1\right>$, where $\mathcal{X}$ is the state space of a system, $\mathcal{A}$ is the  control action (control input) space of an agent, $R:\mathcal{X}\times\mathcal{A}\to\mathbb{R}$ is the immediate reward function, $p_{f}:\mathcal{X}\times\mathcal{X}\times\mathcal{A}\to\mathbb{R}_{\ge0}$ is the probability density of state transitions, and $\rho_1:\mathcal{X}\to\mathbb{R}_{\ge0}$ is the probability density of an initial state. The objective of RL is to learn a control policy $\mu$ that
achieves high cumulative rewards through interactions with the system. To evaluate a policy, the value function $V^{\mu}(x)$ and the action-value function (Q-function) $Q^{\mu}(x,a)$ are commonly used \citep{Sutton_RL}, which represent the expected cumulative rewards starting from state $x$ and state-action pair $(x,a)$, respectively. When the state and action spaces are continuous, {\it actor-critic} methods are commonly employed, where an {\it actor} represents a
control policy, while a {\it critic} estimates a value
function or an action-value function for policy evaluation. The DDPG algorithm is a representative actor-critic algorithm \citep{DDPG}. 

\subsection{Deep Deterministic Policy Gradient (DDPG)}
The DDPG algorithm is a representative actor-critic algorithm for continuous control problems. In this algorithm, we employ two types of DNNs: the {\it critic DNN} $Q_{\theta_{Q}}:\mathcal{X}\times\mathcal{A}\to\mathbb{R}$ and {\it actor DNN} $\mu_{\theta_{\mu}}:\mathcal{X}\to\mathcal{A}$, where $\theta_{Q}$ and $\theta_{\mu}$ denote the parameter vectors of the critic DNN and the actor DNN, respectively.

At each time $t=1,2,...$, the agent observes the state of the system $x_t$ and determines the control action $\mu_{\theta_{\mu}}(x_t)$. The agent then injects a noise $\epsilon_t$, which is drawn from an arbitrary stochastic process, into the control action $a_t=\mu_{\theta_{\mu}}(x_t)+\epsilon_t$ for exploration. By executing $a_t$, the agent receives the next state $x_{t+1}$ and the corresponding immediate reward $r_t=R(x_t,a_t)$. The agent stores the experience $e_t=(x_t,a_t,x_{t+1},r_{t})$ in the buffer $\mathcal{D}$, which is called the {\it replay buffer}. In parallel with explorations, the agent samples the $N$ experiences $e^{(n)}=(x^{(n)},a^{(n)},x'^{(n)},r^{(n)}),\ n=1,2,...,N$ from $\mathcal{D}$ randomly and updates the parameter vectors $\theta_{Q}$ and $\theta_{\mu}$ using the experiences, which is called the {\it experience replay}. The technique helps to prevent the agent from learning from data that are temporally correlated or biased \citep{DQN}. The parameter vector of the critic DNN $\theta_{Q}$ is updated by minimizing the following \it temporal difference error\rm.
\begin{eqnarray}
&&L=\frac{1}{N}\sum_{n=1}^{N}(y^{(n)}-Q_{\theta_{Q}}(x^{(n)},a^{(n)}))^{2},\label{td_error}\\
&&y^{(n)}=r^{(n)}+\gamma Q_{\theta_{Q}^{-}}(x'^{(n)},\mu_{\theta_{\mu}^{-}}(x'^{(n)})),\label{td_target}
\end{eqnarray}
where target values $y^{(n)}$ are generated by the {\it target critic DNN} $Q_{\theta_{Q}^{-}}$ and the {\it target actor DNN} $\mu_{\theta_{\mu}^{-}}$, rather than the original critic and actor DNNs. 
Since the target DNNs are updated more slowly than the original networks, the temporal-difference targets vary more smoothly during training, which helps stabilize the learning process \citep{DQN}.
The parameter vector of the actor DNN $\theta_{\mu}$ is updated using the following policy gradient.  
\begin{eqnarray}
&&\nabla_{\theta_{\mu}}J(\theta_{\mu})\simeq\nonumber\\
&&\frac{1}{N}\sum_{n=1}^{N}\nabla_{a}Q_{\theta_{Q}}(x,a)|_{x=x^{(n)},a=\mu_{\theta}(x^{(n)})}\nabla_{\theta_{\mu}}\mu_{\theta_{\mu}}(x)|_{x=x^{(n)}}.\nonumber\\ \label{PG}
\end{eqnarray}
The parameter vectors of the target DNNs are updated as follows: $\theta_{Q}^{-}\leftarrow\kappa\theta_{Q}+(1-\kappa)\theta_{Q}^{-},\ \theta_{\mu}^{-}\leftarrow\kappa\theta_{\mu}+(1-\kappa)\theta_{\mu}^{-},$
where $\kappa\in(0,1)$. When $\kappa\ll 1$, the target DNNs can be updated slowly. In general, the above exploration and learning process is repeated with periodic reinitialization of the system state. One such trial-and-error sequence is referred to as an {\it episode}.

\subsection{Domain Randomization (DR)}
When DRL is applied to controller design for physical systems, training is often conducted in simulators to improve safety and reduce operational costs. To mitigate the reality gap, DR is useful \citep{DR,understanding_DR}. In DR, system parameters are randomly varied during training so that a single policy becomes robust across a predefined parameter range $\Xi$. Typically, a parameter configuration is sampled at the beginning of each training episode and fixed during the episode.

However, when the parameter space is large and system behaviors vary significantly across parameters, learning a uniformly effective policy over the entire parameter space can become challenging. Furthermore, when system parameters are not directly observable, the control policy may need to infer the latent parameter information from observation histories. 
Rather than learning a single policy over the entire parameter space $\Xi$, we prepare multiple expert policies to alleviate the difficulty of learning under large parameter variations, and switch among them online through a lightweight adaptation mechanism.

\section{Proposed Method}
\begin{figure*}
\begin{center}
\includegraphics[width=15.5cm]{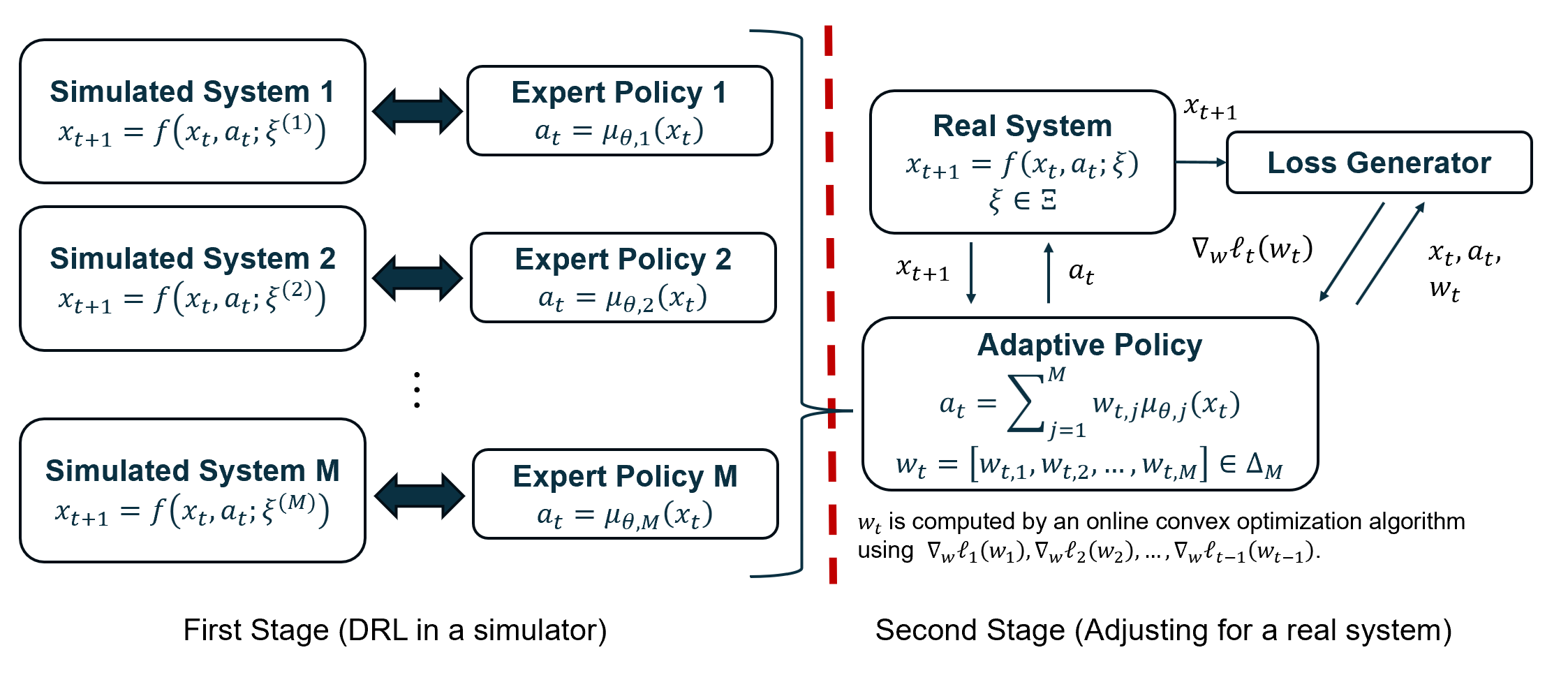}    
\caption{Illustration of the proposed two-stage algorithm, which is called the soft switching expert policies algorithm. } 
\label{fig:SSEP}
\end{center}
\end{figure*}

We propose the two-stage algorithm shown in Fig. \ref{fig:SSEP} for controlling nonlinear systems with uncertain system parameters using DRL and a simulator. In the first stage, multiple representative points are selected from the system parameter space $\Xi$ and a control policy is learned for each corresponding simulated system, which is referred to as an {\it expert policy}. In the second stage, the expert policies are combined through a convex combination, where the weights are adjusted online based on observations from the real system. As a result, the proposed method provides a smooth transition among expert policies. The algorithm is called the \it soft switching expert policies \rm (SSEP) algorithm.

\subsection{Preparing Multiple Expert Policies}
To prepare multiple expert policies, we define the following performance measure of a control policy $\mu$ for a system with a system parameter $\xi\in\Xi$.
\begin{eqnarray}
&&G(\mu|\xi)=\sum_{t=1}^{H}R(x_t,\mu(x_t)),\nonumber\\
&&x_{t+1}=f(x_t,\mu(x_t);\xi),\ x_1=\tilde{x},\nonumber
\end{eqnarray}
where $\tilde{x}$ denotes an initial state given for the evaluation and $H$ denotes the evaluation horizon. We define a control policy $\mu$ as performing poorly on a system with $\xi$ when $G(\mu|\xi)$ is below a given threshold $d$. In this study, several representative points $\{\xi^{(j)}\}_{j=1}^{M}$ are heuristically selected from $\Xi$, and expert policies are learned for the corresponding simulated systems. To improve coverage of the parameter space $\Xi$, we prepare multiple expert policies such that, for any $\xi \in \Xi_{\text{grid}}(\subseteq\Xi)$, at least one
expert policy is expected to achieve acceptable performance on the system with $\xi$, where $\Xi_{\mathrm{grid}}$ denotes a discretized subset of
$\Xi$ used for evaluation. The development of a practical method to automatically select representative points $\{\xi^{(j)}\}_{j=1}^{M}$ remains an important direction for future work.

\noindent {\it Remark:} Note that this finite-horizon measure $G(\mu|\xi)$ is used only for evaluating expert policies. The discount factor $\gamma$ is omitted to evaluate control performance uniformly over the entire trajectory, since discounting may obscure differences in later-stage behavior, particularly in tasks that require long transient behaviors before stabilization (e.g., swing-up pendulum tasks).

\subsection{Adaptive Policy}
We consider the following convex combination of the expert policies learned in a simulator as an adaptive policy.
\begin{eqnarray}
\mu(x,w)=\sum_{j=1}^{M}w_{j}\mu_{\theta_{\mu,j}}(x),\label{adaptive_policy}
\end{eqnarray}
where $w=[w_1,w_2,\dots,w_M]^{\top}$ is an element of the probability simplex $\Delta_M$, i.e., $\sum_{j=1}^{M}w_{j}=1,\ w_{j}\ge0,\ \forall{j}\in\{1,2,\dots,M\}$. Based on the history of observations from the real system, we estimate the similarity vector $w=[w_1,w_2,\dots,w_M]^{\top}\in\Delta_{M}$, which quantifies how accurately each representative simulated system predicts the observed transition of the real system. We then use the vector directly as the weights of the adaptive policy. This design is motivated by the intuition that an expert policy for a representative system is also expected to perform well on systems exhibiting similar behaviors. 

The loss function for the OCO algorithm is defined as follows:
\begin{eqnarray}
\ell_{t}(w_{t})=\left\|x_{t+1}-\sum_{j=1}^{M}w_{t,j}f(x_{t},a_{t};\xi^{(j)})\right\|_{2}^{2},\label{loss_OCO}
\end{eqnarray}
where $x_{t+1}$ results from executing the control action $a_{t}$ in the state $x_{t}$ of the real system. Since $\ell_{t}(w_{t})$ is a squared norm of an affine function of $w_{t}$, it is convex with respect to $w_{t}$. 
In this study, we apply the {\it follow-the-regularized-leader}  (FTRL) algorithm \citep{OCO_text_book}. Let us define
$F(x_t,a_t)
:=
\begin{bmatrix}
f(x_t,a_t;\xi^{(1)}) &
\cdots &
f(x_t,a_t;\xi^{(M)})
\end{bmatrix}
\in\mathbb{R}^{n_x\times M}$. The gradient of the loss $\ell_{t}(w_t)$ with respect to $w_{t}$ is
\begin{eqnarray}
\nabla_{w}\ell_{t}(w_t)=-2F(x_t,a_t)^{\top}(x_{t+1}-F(x_t,a_t)w_t),\nonumber
\end{eqnarray}
which is computed by the {\it loss generator} as shown in Fig. \ref{fig:SSEP}. At each time $t$, the similarity vector is computed by 
\begin{eqnarray}
w_{t}=\arg\min_{w\in\Delta_M} \left\{\sum_{\tau=1}^{t-1}\nabla_{w}\ell_{\tau}(w_{\tau})^{\top}w+\frac{1}{\eta}\Phi(w)\right\},\label{FTRL}
\end{eqnarray}
where $\eta>0$ denotes a learning rate. $\Phi:\Delta_{M}\to\mathbb{R}$ is a regularizer for which we choose \it unnormalized negentropy \rm $
\Phi(w)=\sum_{j=1}^{M}w_{j}\log w_{j}-w_{j}$. Specifically, $w_{t}$ is computed by
\begin{eqnarray}
w_{t,i}=\frac{\exp(-\eta\sum_{\tau=1}^{t-1} \nabla_{w}\ell_{\tau}(w_{\tau})_{i})}{\sum_{j=1}^{M}\exp(-\eta\sum_{\tau=1}^{t-1}\nabla_{w}\ell_{\tau}(w_{\tau})_{j})},\ i\in{1,2,...,M}. \nonumber\\
\label{w_update}
\end{eqnarray}

For convex loss functions, the FTRL algorithm provides sublinear regret guarantees with respect to the best fixed decision in hindsight
\citep{OCO_text_book}. Note that this guarantee concerns the prediction-error loss used for online weight adaptation and does not directly imply closed-loop stability or optimal control performance guarantees.

Various approaches can be considered for adaptively combining expert policies, including {\it mixture-of-experts} architectures \citep{MoE} in which a DNN-based adaptive policy (a gating network) generates the weights of expert policies. However, such DNN-based adaptive policies may require additional retraining or fine-tuning when expert policies are modified or newly introduced. In contrast, the proposed OCO-based approach not only enables lightweight online adaptation but also naturally accommodates the addition or removal of expert policies.

\subsection{Adaptive Policy for Varying System Parameters}
The FTRL algorithm estimates the similarity vector using all past losses, which is suitable when the system parameters of the real system are nearly fixed. However, in general, the system parameters may vary gradually or abruptly due to disturbances. In such cases, losses generated under past system parameters are treated equally, although they may become less relevant to the current system behavior, potentially degrading the estimation of the current similarity vector. To mitigate this effect, we apply the following \it discounted FTRL \rm algorithm.
\begin{eqnarray}
w_{t}=\arg\min_{w\in\Delta_M} \left\{\sum_{\tau=1}^{t-1}\beta^{t-1-\tau}\nabla_{w}\ell_{\tau}(w_{\tau})^{\top}w+\frac{1}{\eta}\Phi(w)\right\},\nonumber\\
\label{discount_FTRL}
\end{eqnarray}
where $\eta$ is a learning rate and $\beta\in(0,1)$ is a discount factor to gradually reduce the effect of losses from the distant past \citep{ADAM_as_FTRL}. Specifically, $w_t$ is computed by 
\begin{eqnarray}
&&w_{t,i}=\frac{\exp(-\eta\sum_{\tau=1}^{t-1}\beta^{t-1-\tau}\nabla_{w}\ell_{\tau}(w_{\tau})_{i})}{\sum_{j=1}^{M}\exp(-\eta\sum_{\tau=1}^{t-1}\beta^{t-1-\tau} \nabla_{w}\ell_{\tau}(w_{\tau})_{j})},\ \nonumber\\
&&\hspace{150pt} i\in{1,2,...,M}. 
\label{w_update_discount}
\end{eqnarray}

\section{Example}
We consider the following discrete-time nonlinear system.
\begin{eqnarray}
\begin{bmatrix}
	x_{t+1,1}\\
	x_{t+1,2}
\end{bmatrix}=\begin{bmatrix}
	x_{t,1}+\delta x_{t,2}\\
	x_{t,2}+\delta\left(\frac{\mathrm{g}}{\mathrm{l}}\sin x_{t,1}-\frac{\xi_{t,2}x_{t,2}}{\xi_{t,1}\mathrm{l}^{2}}+\frac{10.0 a_{t,1}}{\xi_{t,1}\mathrm{l}^2}\right)
\end{bmatrix},\nonumber\\
\label{pendulum}
\end{eqnarray}
where $\mathrm{g}=9.81$, $\delta=0.05$, and $\mathrm{l}=1.0$. The state space is $\mathcal{X}=\mathbb{R}^{2}$, and the control input space is $\mathcal{A}=[-1,1]\subset\mathbb{R}$. Let $\xi=[\xi_{1},\xi_{2}]^{\top}$ denote a vector of system parameters. We assume that $\xi$ is uncertain but lies within the system parameter space $\Xi=\{\xi\in\mathbb{R}^{2}|\ 0.1\le\xi_{1}\le2.0,\ 0.0\le\xi_{2}\le 2.0\}$. The immediate reward function $R$ is defined by 
\begin{eqnarray}
R(x_{t},a_{t})=-x_{t,1}^{2}-0.1x_{t,2}^{2}-10.0a_{t,1}^{2},\label{reward_example}
\end{eqnarray}
that is, our goal is to stabilize the target state $x^{*}=[x_{1}^{*},x_{2}^{*}]^{\top}=[0,0]^{\top}$, which is a fixed point of (\ref{pendulum}). To learn expert policies in a simulator, we apply  DDPG with the same network architectures for actor and critic DNNs. Each DNN consists of two fully connected hidden layers with 128 units per layer. The activation functions of hidden layers are ReLU functions. For the output layers of the actor DNNs, we use hyperbolic tangent functions. The parameter vectors of DNNs are updated using \it Adam \rm \citep{Adam} with learning rates $1.0\times10^{-4}$ for the actor DNNs and $1.0\times 10^{-3}$ for the critic DNNs.

To visualize the robustness of a control policy $\mu$ to discrepancies in system parameters, we plot $G(\mu|\xi)$ for $\xi\in\Xi_{\mathrm{grid}}=\{0.15,0.25,...,1.95\}\times\{0.05,0.15,...,1.95\}$. The initial state for $G(\mu|\xi)$ is set to $\tilde{x}=[\pi\ 0]^{\top}$, the evaluation horizon $H$ is 1000, and the threshold is set to $d=-1500$, i.e., when $G(\mu|\xi)$ is below $-1500$, the policy $\mu$ does not perform well on the system with $\xi$. In all plots of $G(\mu|\xi)$, values below the threshold are shown in black to highlight regions where the policy performs poorly. 

\begin{figure}[h]
    \centering
    \begin{subfigure}{0.45\textwidth}
        \centering
        \includegraphics[width=6.5cm]{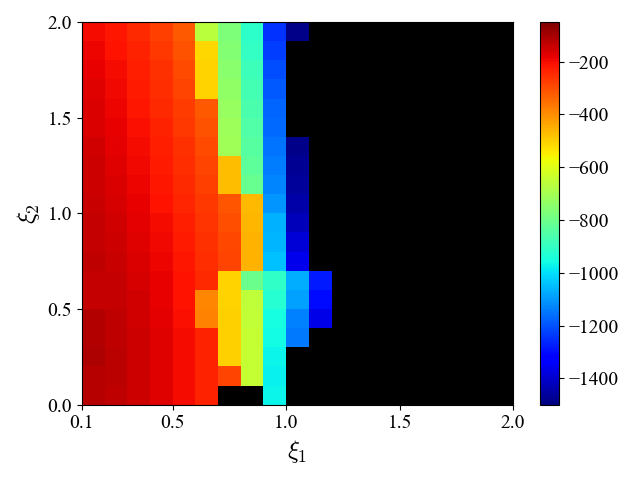}
        \caption{The policy learned without access to the true system parameters}
        \label{fig:DR_without_params}
    \end{subfigure}
    \\
    \begin{subfigure}{0.45\textwidth}
        \centering
        \includegraphics[width=6.5cm]{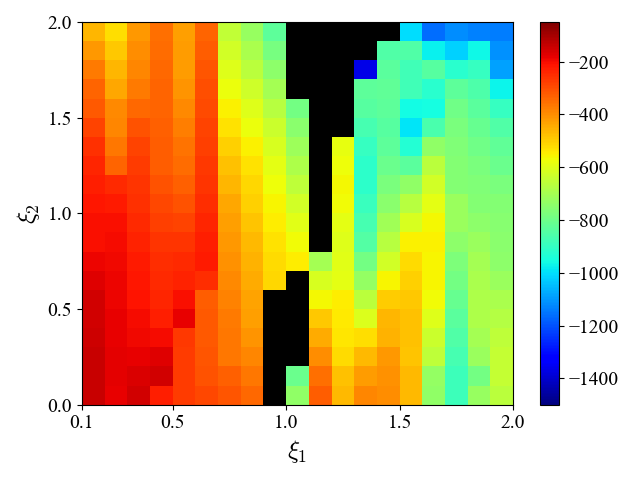}
        \caption{The policy learned with access to the true system parameters}
        \label{fig:DR_with_params}
    \end{subfigure}
    \caption{$G(\mu|\xi)$ of the policies learned using DDPG with DR.}
    \label{fig:DR}
\end{figure}

The performances of the control policies learned using DDPG with DR are shown in Fig. \ref{fig:DR}, where $\xi$ is uniformly sampled from $\Xi$. For simplicity, we do not employ recurrent architectures and instead compare policies trained with and without access to the true system parameters. Fig.~\ref{fig:DR_without_params} shows the performance of the policy learned without access to the true system parameters, whereas Fig.~\ref{fig:DR_with_params} shows that of the policy learned with access to them. These results indicate that it is difficult to learn a robust policy using DR without access to the true system parameters. Additionally, even if the true system parameters are available, the learned policy may not perform well for some parameters in $\Xi$ as shown in Fig.\ \ref{fig:DR_with_params}. The performance may further degrade when the system parameters must be inferred with recurrent architectures.

\subsection{Preparing Expert Policies}
\begin{figure*}[h]
    \centering
    \begin{subfigure}{0.31\textwidth}
        \centering
        \includegraphics[width=5.5cm]{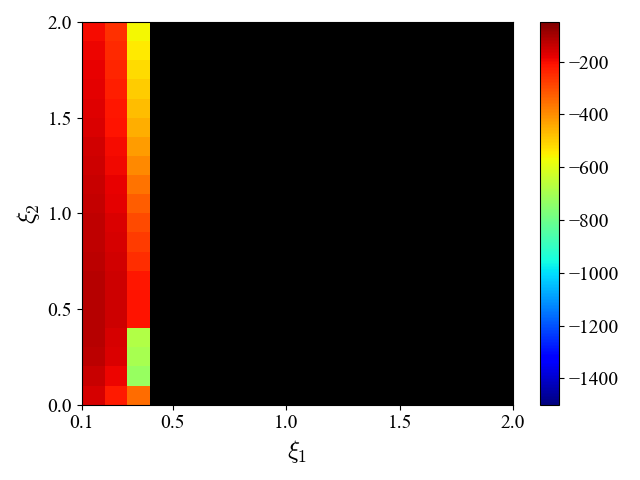}
        \caption{Expert policy learned for $f(\cdot,\cdot;\xi^{(1)})$.}
        \label{fig:expert1}
    \end{subfigure}\hspace{1mm}
    \hfill
    \begin{subfigure}{0.31\textwidth}
        \centering
        \includegraphics[width=5.5cm]{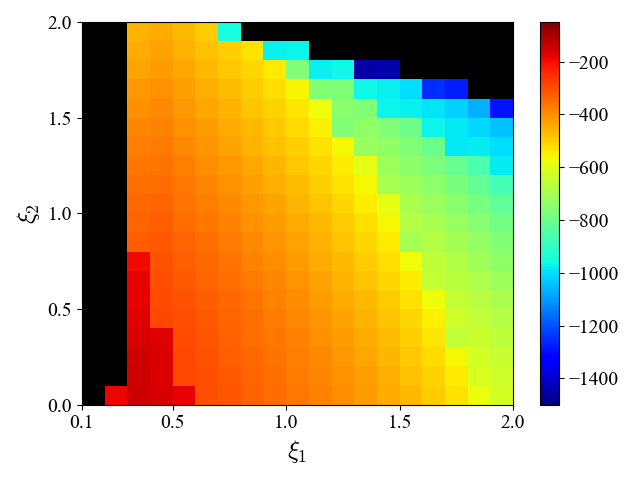}
        \caption{Expert policy learned for $f(\cdot,\cdot;\xi^{(2)})$.}
        \label{fig:expert2}
    \end{subfigure}\hspace{1mm}
    \hfill
    \begin{subfigure}{0.31\textwidth}
        \centering
        \includegraphics[width=5.5cm]{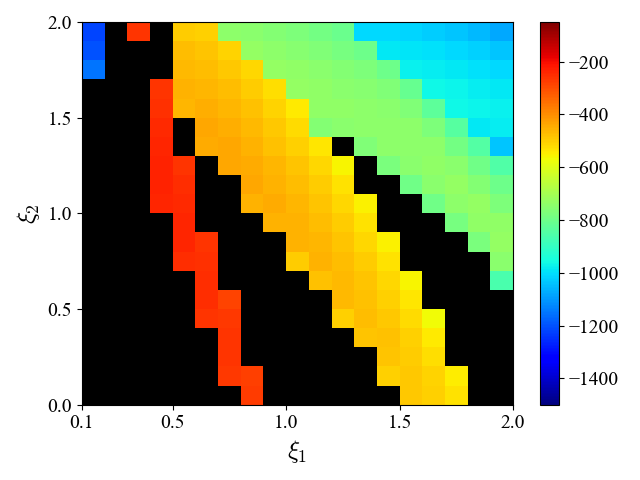}
        \caption{Expert policy learned for $f(\cdot,\cdot;\xi^{(3)})$.}
        \label{fig:expert3}
    \end{subfigure}\hspace{1mm}
    \caption{$G(\mu|\xi)$ of three expert policies learned for representative simulated systems using DDPG.}
    \label{fig:local_experts}
\end{figure*}

The development of a practical method for selecting representative points is beyond the scope of this study; here, we heuristically select representative points from $\Xi$ based on the simulation results $G(\mu|\xi)$ obtained from certain simulated systems $f(\cdot,\cdot,\xi)$. Specifically, we select the following three representative points: $\xi^{(1)}=[0.1,1.0],\ \xi^{(2)}=[2.0, 0.0]$, and $\xi^{(3)}=[2.0,2.0]$. We then synthesize expert policies using DDPG for these representative simulated systems. The performances of the three expert policies are shown in Fig.\ \ref{fig:local_experts}. Each expert policy performs well on systems whose behaviors are similar to those of the corresponding representative simulated system. Additionally, for any $\xi\in\Xi_{\text{grid}}$, there exists at least one expert policy that performs well.

\subsection{Adaptive Policy under Fixed System Parameters}
\begin{figure}
\begin{center}
\includegraphics[width=6.5cm]{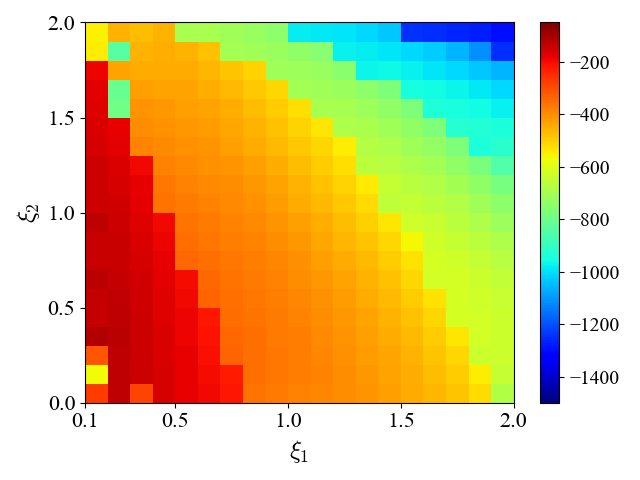}    
\caption{$G(\mu|\xi)$ of the adaptive policy with FTRL.} 
\label{fig:Score_of_SSEP}
\end{center}
\end{figure}

We consider a real system whose system parameters are fixed. The weights of the adaptive policy are adjusted through the estimation of the similarity vector $w$ using the FTRL algorithm every five control steps. We set the learning rate $\eta=0.5$. The performance of the adaptive policy is shown in Fig.\ \ref{fig:Score_of_SSEP}. Using observations from the real system, the adaptive policy achieves performance above the prescribed threshold for all systems in $\Xi_{\text{grid}}$.

\subsection{Adaptive Policy under Varying System Parameters}
\begin{figure}
\begin{center}
\includegraphics[width=8.4cm]{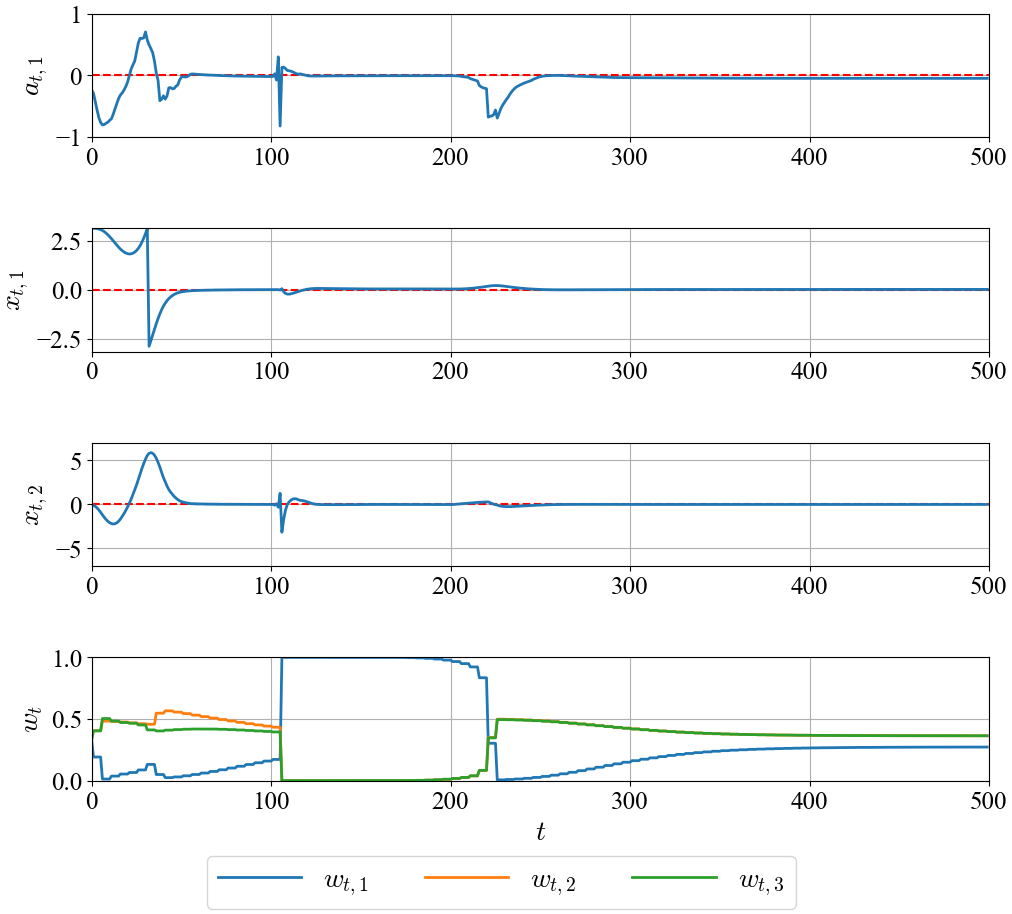}    
\caption{Time response of the system controlled by the adaptive policy with discounted FTRL.} 
\label{fig:TR_of_SSEP}
\end{center}
\end{figure}

We consider a real system whose system parameters vary abruptly several times. We adjust the weights of the adaptive policy using the discounted FTRL algorithm every five control steps. We set the learning rate $\eta=1.0$ and the discount factor $\beta=0.9$. We assume that the system parameters vary as follows:
\begin{eqnarray}
    \xi_{t,1}=\begin{cases}
        1.2 & t\in \mathcal{T}_1\\
        0.1 & t\in \mathcal{T}_2\\
        1.9 & t\in \mathcal{T}_3,
    \end{cases}\ \ \ \ 
    \xi_{t,2}=\begin{cases}
        0.0 & t\in \mathcal{T}_1\\
        0.5 & t\in \mathcal{T}_2\\
        2.0 & t\in \mathcal{T}_3,
    \end{cases}\nonumber
\end{eqnarray}
where these time intervals are $\mathcal{T}_1=[1,100]$, $\mathcal{T}_2=[101,200]$, and $\mathcal{T}_3\in[201,500]$, respectively. 
The time response is shown in Fig.\ \ref{fig:TR_of_SSEP}. The adaptive policy can adjust the weights $w$ in response to varying system parameters. Note that, in the time interval $\mathcal{T}_{3}$, the weights $w^{(2)}$ and $w^{(3)}$ become nearly identical. In this interval, $x_{t,2}$ remains close to zero. Since the effect of $\xi_{t,2}$ appears through the term $-\frac{\xi_{t,2}}{\xi_{t,1}\mathrm{l}^{2}}x_{t,2}$, the influence of $\xi_{t,2}$ on the observed transition becomes less pronounced when $x_{t,2}\simeq 0$. As a result, the behaviors of the representative systems corresponding to $\xi^{(2)}$ and $\xi^{(3)}$ become similar, making it difficult to clearly distinguish them based on observations. In such situations, hard switching schemes may cause chattering behavior, whereas the proposed adaptive policy can smoothly interpolate between expert policies through convex combinations of their actions.

\section{Conclusion}
We proposed a simulation-based two-stage algorithm. First, we select representative systems $f(\cdot,\cdot;\xi^{(j)})$,  $j=1,2,...,M$, and synthesize an expert policy for each representative system using a standard DRL algorithm in a simulator. Second, we construct an adaptive policy as a convex combination of expert policies. The weights are adjusted by an OCO algorithm using observations from a real system. The effectiveness of our proposed algorithm was validated through numerical experiments. An important direction for future research is to develop a method that automatically selects representative systems for preparing expert policies. Another important direction is to establish closed-loop stability or control performance guarantees for the proposed framework. In addition, applying the proposed algorithm to complex real-world systems such as industrial plants is also an important direction.

\bibliography{ifacconf}             

\end{document}